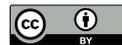

*Apuntes* 95, 171-202  
ISSN: 0252-1865  
eISSN: 2223-1757  
doi: 10.21678/apuntes.95.1811

© Creative Commons Attribution 3.0  
Artículo recibido el 18 de marzo de 2022  
Artículo aceptado el 5 de julio de 2023# ¿Es posible obtener estimaciones confiables del porcentaje de anemia y retraso en el crecimiento en niños menores de cinco años en los distritos más pobres del Perú?

Anna Sikov  
*Universidad Nacional de Ingeniería, Perú*  
asikov@uni.edu.pe

José Javier Cerda Hernández  
*Universidad Nacional de Ingeniería, Perú*  
jcerdah@uni.edu.pe

Marcial Eduardo Haro Abanto  
*Pontificia Universidad Católica del Perú*  
eduardo.haro@pucp.edu.pe*Resumen.* En este artículo, obtenemos predicciones confiables del porcentaje de niños con anemia y el porcentaje de niños con retraso del crecimiento por distrito en el Perú, utilizando los datos de la Endes del año 2019 y del censo nacional realizado el año 2017, en los distritos donde el tamaño de la muestra no es suficiente para implementar una estimación directa, y en los distritos no muestreados. Como el objetivo principal de las encuestas nacionales es describir el estado de la población (por ejemplo, la salud, el estado de empleo y desempleo, gastos familiares, educación, etc.), uno de los problemas más comunes de las encuestas nacionales es que estas son generalmente planeadas de tal forma que tengan una buena representación solamente a nivel nacional, nacional urbano, nacional rural o región natural. Por tal motivo, la inferencia a niveles más desagregados, como a nivel distrital o provincial, no es confiable por tener muestras pequeñas a estos niveles. Es decir, hay muchos distritos que no fueron incluidos en la muestra o no cuentan con suficientes observaciones para realizar estimaciones adecuadas al nivel provincial y distrital utilizando estimadores directos. En este trabajo, presentamos y aplicamos el modelo de Fay-Herriot espacial (Pratesi, M., & Salvati, N. (2008)) para obtener estimaciones robustas de la prevalencia de anemia y de retraso en el crecimiento en la niñez en los distritos del Perú donde no se tiene observaciones o estas son pocas para poder hacer inferencia. Este tipo de modelos usa la información global del censo y la combina con la información local y espacial de la Endes para obtener estimadores fiables de las variables de interés.

*Palabras clave.* Anemia en niños, retraso en el crecimiento de niños, desnutrición infantil, estimador directo, autocorrelación espacial, modelo de Fay-Herriot, modelo de Fay-Herriot espacial.

**Is it possible to obtain reliable estimates for the prevalence of anemia and childhood stunting among children under 5 in the poorest districts in Peru?**

*Abstract.* In this article we describe and apply the Fay-Herriot model with spatially correlated random area effects (Pratesi, M., & Salvati, N. (2008)), in order to predict the prevalence of anemia and childhood stunting in Peruvian districts, based on the data from the Demographic and Family Health Survey of the year 2019, which collects data about anemia and childhood stunting for children under the age of 12 years, and the National Census carried out in 2017. Our main objective is to produce reliable predictions for the districts, where sample sizes are too small to provide good direct estimates, and for the districts, which were not included in the sample. The basic Fay-Herriot model (Fay & Herriot, 1979) tackles this problem by incorporating auxiliary information, which is generally available from administrative or census records. The Fay-Herriot model with spatially correlated random area effects, in addition to auxiliary information, incorporates geographic information about the areas, such as latitude and longitude. This permits modeling spatial autocorrelations, which are not unusual in socioeconomic and health surveys. To evaluate the mean square error of the above-mentioned predictors, we use the parametric bootstrap procedure, developed in Molina et al. (2009).

¿Es posible obtener estimaciones confiables del porcentaje de anemia y retraso en el crecimiento en niños menores de cinco años en los distritos más pobres del Perú?

## 1. Introducción

Según la Organización Mundial de la Salud (OMS), la anemia en la niñez es un trastorno en la capacidad de transporte de oxígeno en la sangre, lo cual dificulta satisfacer las necesidades del organismo y tiene efectos como fatiga, debilidad y retraso en el crecimiento. A su vez, el retraso en el crecimiento en la niñez, producto de la prevalencia de desnutrición crónica, se define como una estatura inferior en más de dos desviaciones estándar a la mediana de los patrones de crecimiento infantil de la OMS y tiene efectos a largo plazo, como mala salud y disminución de la capacidad cognitiva, entre otros (Beard & Connor, 2003; Buttenheim, Alderman, & Friedman, 2011; World Health Organization. Centers for Disease Control and Prevention, 2004). Por tal motivo, la reducción de la anemia y la desnutrición infantil, que se ven reflejadas en el retraso en el crecimiento, es una prioridad en las políticas de salud del Estado peruano. Según el Plan Nacional para la Reducción y Control de la Anemia Materno Infantil y la Desnutrición Crónica Infantil en el Perú: 2017-2021, del Ministerio de Salud, la meta para el Bicentenario sobre el porcentaje de anemia infantil era del 19%. Sin embargo, la cifra reportada por el Minsa en 2018 fue del 43,5%, consiguiendo reducir solamente un 3,3%, desde el inicio del primer plan nacional contra la anemia de 2014. Según Francke y Acosta (2020), a este ritmo de reducción, «el objetivo de 19% no se lograría sino hasta el 2048». Frente a esta situación, el Perú viene implementando diversos programas sociales, como el Vaso de Leche, Juntos y Qali Warma, con el objetivo de reducir la anemia y la desnutrición infantil. Lo más importante de estos programas es cuantificar el impacto que tienen estos programas sobre la reducción de la anemia y la desnutrición, puesto que los costos son muy altos para el Perú: solo para el caso de la reducción de la anemia, representan alrededor del 0,62% del producto interno bruto del país (para más detalles sobre este cálculo, véase Alcázar, 2012). Además, según el Midis, el presupuesto asignado para el programa Qali Warma para 2022 es de S/ 1987 millones.

En este artículo, analizamos y predecimos el porcentaje de niños menores de cinco años con anemia y el porcentaje de niños menores de cinco años con riesgo de retraso en el crecimiento por distrito, utilizando los datos de la Endes del año 2019 y del censo nacional realizado el año 2017. En este trabajo, definimos como un niño con riesgo de retraso en el crecimiento a aquel cuya su altura es menor que el valor que corresponde al percentil 10 según la distribución de altura dada por la OMS correspondiente a su edad. Según la Endes 2019, el promedio nacional de prevalencia de anemia en niños menores de 36 meses es del 40,1%, y esta llega a valores entre 50 y 70% en departamentos como Puno, Cusco, Huancavelica y Ucayali. Sin embargo, estas cifras no reflejan el verdadero valor de la prevalencia de anemia en los niños, puesto que el muestreo usado por el INEI para la Endes, por cuestiones logísticas y de restricciones presupuestarias, no toma en cuenta muchos distritos del Perú donde generalmente la evidencia empírica demuestra que la prevalencia de anemia es mayor. El objetivo del presente artículo es estimar el porcentaje de niños con anemia y el porcentaje de niños con riesgo de retraso en el crecimiento en los distritos donde no se tiene observaciones o se tiene pocas observaciones, usando una metodología estadística con la que se construyen estimadores insesgados de los parámetros del modelo utilizado. Estas estimaciones ayudarían a implementar más eficientemente los programas sociales en el país, dirigiendo más recursos a las regiones con mayor prevalencia de anemia y retraso en el crecimiento o menor impacto de los programas sociales, lo que daría a los implementadores de políticas públicas un mejor mapa de anemia y retraso en el crecimiento del Perú para la toma de decisiones y distribución de recursos.

El problema de muestreo descrito anteriormente se genera porque el objetivo principal de las encuestas nacionales es describir el estado de la población (por ejemplo, la salud, el estado de empleo y desempleo, gastos familiares, educación, etc.), y uno de los problemas más comunes de las encuestas nacionales es que estas son generalmente planeadas de tal forma que tengan una buena representación solo a nivel nacional, nacional urbano, nacional rural o región natural. Por tal motivo, la inferencia directa a partir de los datos de la Endes u otras encuestas del INEI a niveles más desagregados no es confiable, puesto que a estos niveles la muestra por distrito o provincia es muy pequeña o nula. Es decir, hay muchos distritos que no fueron incluidos en la muestra o no cuentan con suficientes observaciones para realizar estimaciones adecuadas a nivel provincial o distrital utilizando estimadores directos. Por ejemplo, en el departamento de Puno solo el 41,4% de distritos tienen datos de prevalencia de anemia, siendo los distritos de la provincia de Juliaca y de Puno aquellos que presentan un mayor número de encuestados (más de 70 observaciones), en tanto que el resto de los distritos tienen pocas observaciones o no fueron muestreados. Según la Endes de 2019, la prevalencia de anemia en el departamento de Puno es del 69,9%, pero este porcentaje podría ser mayor si incluimos al 58,6% de los distritos excluidos en la muestra de dicha encuesta para dicho departamento. Nuestro objetivo principal es obtener predicciones confiables y robustas en los distritos donde el tamaño de la muestra no es suficiente para implementar una estimación directa, y en los distritos no muestreados. Estas estimaciones darían una mejor información a los hacedores de políticas públicas sobre el verdadero mapa de anemia del Perú, y ayudarían a mejorar, a través de pruebas piloto, el muestreo realizado por el INEI para obtener una muestra más representativa a diferentes niveles y describir mejor diversas variables socioeconómicas del Perú.

Para abordar este problema, entidades gubernamentales como la Oficina Europea de Estadística, la Oficina del Censo de los Estados Unidos entre muchos otros, utilizan el modelo de Fay-Herriot (Fay & Herriot, 1979), que es un modelo a nivel de área (a un nivel más desagregado, que en nuestro caso sería por distrito) que constituye una combinación de predicción basada en la estimación directa bajo el diseño de muestreo y el modelo de regresión usando variables exógenas dadas por el censo. El modelo de regresión incorpora la información auxiliar a nivel del área, que generalmente está disponible en las fuentes como censos o registros administrativos gubernamentales, etc. El modelo de Fay-Herriot espacial, además, incorpora la información sobre las localizaciones geográficas de las áreas (los distritos en este estudio), lo que permite tomar en cuenta las autocorrelaciones espaciales, que comúnmente presentan los datos de encuestas socioeconómicas. Generalmente, los datos de la Endes muestran que distritos vecinos pobres tienen un nivel de prevalencia de anemia similar.

De esta manera, por un lado, las áreas (distritos) son enlazadas mediante el vector de los coeficientes de regresión y, por otro lado, cada área tiene su efecto aleatorio, que recoge la variación que no está explicada por las variables auxiliares. El modelo de Fay-Herriot básico supone que los efectos aleatorios son independientes e idénticamente distribuidos. Para realizar predicciones en las áreas que tienen un número pequeño de observaciones y en las áreas no muestreadas, los autores desarrollan el mejor predictor lineal insesgado (BLUP, por sus siglas en inglés: *best linear unbaised predictor*). Dicho predictor constituye un estimador compuesto, que se obtiene ponderando el estimador directo de la encuesta y el estimador sintético de regresión, y supone que los elementos de la matriz de las covarianzas que corresponde a la componente aleatoria del modelo, son conocidos. En la realidad, la varianza de los efectos aleatorios del modelo es desconocida y, por lo tanto, se estima utilizando el método máxima verosimilitud, el método de máxima verosimilitud restringida o el método de los momentos. Reemplazando los parámetros desconocidos en el predictor BLUP por sus estimadores respectivos, se obtiene el mejor predictor lineal insesgado empírico Eblup (*empirical best linear unbiased predictor*).

El estudio previo de los datos de la Endes demuestra que ellos presentan autocorrelaciones muy fuertes a lo largo del espacio (véase Moran, 1950), en el sentido de que la información espacial de un distrito es generalmente muy relevante para poder predecir el valor de la variable de interés en el distrito vecino. Por ejemplo, si en un distrito el porcentaje de niños con retraso en el crecimiento es alto, es muy probable que el mismo fenómeno sea observado en los distritos vecinos. Para resolver el problema de la presencia de autocorrelaciones espaciales en la variable de interés, se puede utilizar una extensión del modelo de Fay-Herriot, el modelo de Fay-Herriot espacial (véanse, para más detalles, Singh, 2005; Petrucci & Salvati, 2006; Pratesi & Salvati, 2008). Dicho modelo permite incorporar las autocorrelaciones espaciales modelando los efectos aleatorios mencionados arriba mediante el proceso SAR (*simulataneously autoregressive model*), que se caracteriza por un parámetro autorregresivo y una matriz de vecindad, que define la proximidad entre cada par de áreas (distritos en nuestro caso) (Anselin, 1992; Cressie, 1993). De esta manera, el efecto aleatorio de un área específica es una combinación lineal de los efectos aleatorios de las áreas vecinas. Para obtener las predicciones bajo el modelo de Fay-Herriot espacial, se utiliza el predictor SBLUP (*spatial best linear unbiased predictor*), propuesto en Pratesi & Salvati (2008). Reemplazando los parámetros desconocidos por los estimadores correspondientes, el autor obtiene el SBLUP empírico Seblup.

En este artículo, aplicamos el modelo de Fay-Herriot y el modelo de Fay-Herriot espacial para predecir el porcentaje de niños anémicos y el porcentaje de niños con riesgo de retraso en el crecimiento por distrito, utilizando los datos de la Encuesta Demográfica de Salud Familiar (Endes) y los datos del censo nacional del año 2017. Como ya fue mencionado, los predictores Eblup bajo el modelo Fay-Herriot y Seblup bajo el modelo Fay-Herriot espacial son más fiables que los estimadores directos, especialmente para los distritos con pocas observaciones. Además, al contrario de los estimadores directos, los predictores Eblup y Seblup pueden ser utilizados para predecir los porcentajes de niños anémicos y de niños con riesgo de retraso en el crecimiento en los distritos que no fueron seleccionados en la muestra por el INEI.

## 2. Exploración empírica de los datos y marco metodológico

### 2.1 Base de datos y justificación

Para el presente trabajo, la principal fuente de información es la Encuesta Demográfica y de Salud Familiar (Endes) del año 2019 y el censo nacional realizado el año 2017. En el caso de la Endes, esta es una de las encuestas anuales elaboradas por el Instituto Nacional de Estadística e Informática (INEI), que tiene como objetivo principal recopilar información de indicadores nutricionales y de salud de los niños menores de cinco años nacidos de mujeres con edades entre 15 y 49 años, así como obtener información detallada de las características del hogar y de la madre. Entre las principales estadísticas que se derivan de esta encuesta, se encuentran la tasa de fecundidad, el uso de métodos de planificación familiar, la prevalencia de anemia en niños, entre otros. La Endes es una encuesta bietápica a nivel nacional, cuyas unidades de muestreo son viviendas. Dicha encuesta se realiza de tal manera que, en la primera etapa, se seleccionan conglomerados en las áreas urbanas y áreas de empadronamiento rural en las áreas rurales, en todos los departamentos del país con probabilidad proporcional a su tamaño, empleando la información del Censo de Población y Vivienda del 2007 y el Sistema de Focalización de Hogares 2012-2013. En la segunda etapa, se eligieron viviendas de los conglomerados y áreas de empadronamiento rural seleccionados anteriormente de tal manera que la muestra sea equilibrada considerando las variables: niñas y niños menores de cinco años, mujeres en edad fértil, etc. (véase INEI, 2019). La Endes tiene una buena representación a nivel nacional y departamental. Por ejemplo, en el año 2019, el número total de viviendas muestreadas fue 36 760. Sin embargo, la inferencia directa a niveles más desagregados, como provincial y distrital, no es confiable, puesto que hay muchos distritos que no fueron incluidos en la muestra o no cuentan con suficientes observaciones para realizar estimaciones adecuadas a estos niveles utilizando estimadores directos. La tabla 2.1 muestra el número de distritos no considerados en la Endes por cada departamento. Por ejemplo, en el departamento de Puno, solo el 41,4% de los distritos tienen datos de prevalencia de anemia, siendo los distritos de la provincia de Juliaca y de Puno aquellos que presentan un mayor número de encuestados (más de 70 observaciones), en tanto que el resto de los distritos tienen pocas observaciones o no fueron muestreados. Según la Endes de 2019, la prevalencia de anemia en el departamento de Puno es del 69,9%, pero este porcentaje podría ser mayor si consideramos todos los distritos que pertenecen a dicho departamento.

Para el caso del censo nacional, este es un procedimiento de carácter estadístico mediante el cual el Gobierno peruano recolecta información total de su población, y cuyos datos son empleados luego para la implementación de políticas públicas. Por el costo y la logística que involucra un censo, la Ley N.º 13248, Ley de Censos, considerada como Ley Orgánica de los Censos en el Perú, dispone que, a partir de 1960, los Censos Nacionales de Población y Vivienda deberán realizarse cada 10 años. El organismo público encargado de esta actividad es el INEI. El último censo fue realizado en 2017. En el presente trabajo, usaremos el censo nacional de 2017 para seleccionar un *set* de variables exógenas para incorporarlas como información auxiliar a nivel de distrito. Además, usaremos la información sobre las localizaciones geográficas de los distritos en este estudio para tomar en cuenta las autoco-rrelaciones espaciales, que comúnmente presentan los datos de encuestas socioeconómicas, e incorporar dicha información espacial para tener una mejor estimación de la prevalencia de anemia y el riesgo de retraso en el crecimiento en distritos donde se tienen pocas observaciones o ninguna. Del censo nacional se seleccionarán variables que expliquen el porcentaje de anemia y el retraso en el crecimiento en niños menores de cinco años en los distritos del Perú.

### 2.2 Modelo econométrico para estimar el porcentaje de anemia y retraso en el crecimiento en distritos no muestreados o con pocas observaciones

Como fue reportado en Alcázar (2012), los costos en educación, salud y desarrollo integral del niño representa alrededor del 0,62% del PIB, y según el presupuesto anual de 2021, el Perú destinó S/ 1987 millones solo para el programa Qali Warma, pero la cobertura aún no llega al 100%. Este tipo de programas sociales tiene como uno de sus principales objetivos reducir la anemia en los niños menores de cinco años por el efecto negativo que tiene este a largo plazo en la salud y la disminución de la capacidad cognitiva, limitando su desarrollo como individuo en la sociedad (Beard & Connor, 2003; Buttenheim *et al.*, 2011; World Health Organization. Centers for Disease Control and Prevention, 2004) e impactando directamente en la economía del país (Alcázar, 2012; Horton & Ross, 1998; Martínez & Fernández, 2009). Entonces, el primer problema que el Estado peruano tiene que resolver a traves de políticas públicas es el monitoreo continuo de la población vulnerable al problema de la anemia y la desnutrición infantil. Este monitoreo es parcialmente resuelto a través de la Endes (y otras encuestas), que recopila información de indicadores nutricionales y de salud de los niños menores de cinco años, pero solo es de una muestra de la población objetivo. Para la mejor implementación de los programas sociales dirigidos a reducir la anemia, Francke y Acosta (2020) señalan que los programas sociales «se organizan bajo un modelo de cogestión, que por su naturaleza, necesita la participación y cooperación de diferentes actores de la sociedad civil, tanto del sector público como del privado"» (p. 156), pero el principal problema de la Endes es que la inferencia confiable e informativa que se puede obtener de ella es solo a nivel nacional o departamental, no serviría para administraciones públicas menores, como provincias y distritos.

En el diseño del muestreo para la ENDES, muchos distritos no son considerados en la muestra por diversos factores, o tienen pocas observaciones para poder hacer una inferencia confiable sobre la anemia y el riesgo de retraso en el crecimiento con estimadores directos. En la tabla 2.1, mostramos el porcentaje de distritos excluidos para la Endes 2019. Para más detalle sobre la construcción de la Endes, el lector puede revisar INEI (2019). En el caso de los distritos que no forman parte de la muestra, los gobiernos locales no tendrían información local para abordar el problema de la anemia o desnutrición en forma colaborativa en su comunidad, mientras que los distritos con pocas observaciones tendrían estimaciones sesgadas del porcentaje de anemia y retraso en el crecimiento, generadas por una varianza grande del estimador como consecuencia de la muestra pequeña. En la figura 2.1, construida a partir de la Endes 2019, mostramos los departamentos de La Libertad y Arequipa, donde podemos observar que muchos de los distritos que forman parte de la muestra tienen menos de 20 observaciones. Cabe notar también que, en el caso de los departamentos La Libertad y Arequipa, el 57,8% y el 69,7% de los distritos no fueron incluidos en la muestra. Véase la tabla 2.1.

Tabla 2.1
Número total de distritos por departamento

| Departamento | Total de distritos | N.º de distritos no encuestados | % | Población en distritos no encuestados | % |
|---|---|---|---|---|---|
| Amazonas | 84 | 41 | 49,8 | 42 328 | 11,9 |
| Áncash | 166 | 129 | 77,7 | 511 421 | 49,6 |
| Apurímac | 84 | 47 | 56,0 | 87 117 | 22,6 |
| Arequipa | 109 | 76 | 69,7 | 182 467 | 13,8 |
| Ayacucho | 119 | 76 | 63,9 | 148 615 | 25,4 |
| Cajamarca | 127 | 83 | 65,4 | 428 931 | 33,8 |
| Callao | 7 | 1 | 14,3 | 3745 | 0,4 |
| Cusco | 112 | 71 | 63,4 | 443 402 | 38,7 |
| Huancavelica | 100 | 62 | 62,0 | 112 017 | 34,0 |
| Huánuco | 84 | 45 | 53,4 | 233 550 | 34,3 |
| Ica | 43 | 18 | 41,9 | 102 414 | 12,7 |
| Junín | 124 | 80 | 64,5 | 324 922 | 27,5 |
| La Libertad | 83 | 48 | 57,8 | 309 260 | 18,4 |
| Lambayeque | 38 | 8 | 21,1 | 54 866 | 4,8 |
| Lima | 171 | 109 | 63,7 | 2 302 501 | 25,3 |
| Loreto | 53 | 21 | 39,6 | 108 517 | 13,1 |
| Madre de Dios | 11 | 0 | 0,0 | 0 | 0,0 |
| Moquegua | 20 | 9 | 45,0 | 15 628 | 9,3 |
| Pasco | 29 | 4 | 13,8 | 9915 | 4,1 |
| Piura | 65 | 31 | 47,7 | 349 055 | 19,9 |
| Puno | 110 | 72 | 65,5 | 397 148 | 35,4 |
| San Martín | 77 | 36 | 46,8 | 175 408 | 22,9 |
| Tacna | 28 | 16 | 57,1 | 123 560 | 39,1 |
| Tumbes | 13 | 3 | 23,1 | 36 864 | 17,4 |
| Ucayali | 17 | 5 | 29,4 | 24 522 | 5,3 |

Notas. La tercera y la cuarta columna nos muestran el número de distritos no considerados en la muestra de la Endes 2019 y el porcentaje que representa este del total de distritos, respectivamente. La quinta y la sexta columna nos muestran el tamaño total de la población objetivo en los distritos no considerados en la Endes y el porcentaje que representa del total de distritos, respectivamente.
Fuente: elaboración propia con base en los datos de la Endes 2019.

Figura 2.1
Tamaño de la muestra en cada uno de los distritos seleccionados para la Endes 2019

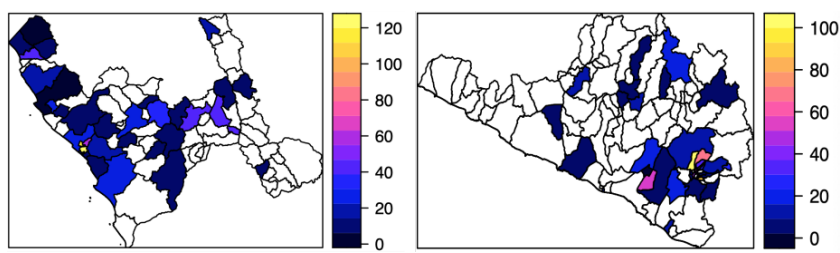

Nota. Esta figura muestra el tamaño de la muestra en cada uno de los distritos seleccionados para la Endes 2019 (lado izquierdo, La Libertad; lado derecho, Arequipa). Los distritos en color blanco no fueron seleccionados para el muestreo.
Fuente: elaboración propia con base en los datos de la ENDES 2019.

La pregunta que se busca responder en el presente trabajo es la siguiente: ¿es posible obtener estimaciones confiables del porcentaje de anemia y retraso en el crecimiento en niños menores de cinco años en distritos que no forman parte de la muestra de la Endes o donde se tiene pocas observaciones?

Este estudio pretende dar un soporte al muestreo diseñado por el INEI para sus diferentes encuestas y, con ello, ayudar a mejorar los procesos de selección de la muestra, para así obtener mejores estimaciones de las variables socioeconómicas que usan los diversos organismos públicos para el diseño de políticas públicas. La idea general de los métodos presentados en este trabajo es combinar diversas fuentes de información e incluirlas en un modelo que capture diversas características sociales, económicas y regionales de la variable de interés a nivel de distritos. El presente trabajo aplica métodos cuantitativos del tipo econométrico basados en los modelos de Fay y Herriot (1979) simple y espacial, con el objetivo de estimar las variables de interés en los distritos no muestreados. En la siguiente sección, describimos la metodología propuesta para estimar dichos porcentajes.

### 2.2.1 Modelo de Fay-Herriot

Denotemos por $Y_i$ el valor del estimador directo de la característica o variable de interés (por ejemplo, para el presente trabajo, $Y_i$ representa el porcentaje de los niños con anemia o con riesgo de retraso en el crecimiento) en el distrito $i$, $i = 1, ..., D$, donde $D$ es el número total de los distritos para los cuales hay datos disponibles en la encuesta; y, por $\theta_i$, el verdadero valor de la misma característica o variable de interés en el mismo distrito. La característica $Y_i$ es tomada de la Endes. Además, denotemos por $X_i$ el vector de las variables auxiliares o exógenas (por ejemplo, el porcentaje de la población que tiene acceso al agua potable dentro de la vivienda, el porcentaje de la población que no tiene ningún tipo de seguro de salud, etc.). En nuestro caso, los valores de las variables auxiliares provienen del Censo Nacional de 2017, y por lo tanto, no tienen errores de medida y, además, son observados para todos los distritos del Perú. Este tipo de uso de variables auxiliares de otras fuentes de información es conocida como *borrow information*.

El modelo de Fay-Herriot asume que el valor observado de la característica de interés en un distrito es una perturbación normal de media cero del verdadero valor de la misma característica, y que, a su vez, el verdadero valor es una perturbación normal de media cero de una combinación lineal de un *set* de variables explicativas que no tienen errores de medida. Estas variables vendrían de fuentes como los censos. Así, el modelo de Fay-Herriot con variables auxiliares sin errores de medida se define de la siguiente forma:

$$Y_i = \theta_i + e_i; \qquad \theta_i = X_i\beta + u_i, \quad (2.1)$$

donde $e_i \sim N(0,\sigma_i^2)$ son los errores del estimador directo y $u_i \sim N(0,\sigma_u^2)$ son efectos aleatorios, que representan la heterogeneidad entre las áreas (distritos) no explicada por las variables auxiliares; y $\beta$ es el vector de los coeficientes que refleja la asociación entre $\theta = (\theta_1,...,\theta_D)^t$ y $X = (X_1,...,X_D)^t$. Se supone que los errores $e_i$ y los efectos aleatorios $u_i$ para $i = 1, ..., D$ son independientes de los otros distritos, es decir, $cov(e_i,e_j) = cov(u_i,u_j) = 0$ si $i \neq j$ y, además, los errores de muestreo y los efectos aleatorios también son independientes, tal que $cov(e_i,u_j) = 0 \ \forall \ i,j$.

Observen que el vector de los coeficientes $\beta$ no depende del distrito, es decir, el impacto de las variables $X_i$ sobre el verdadero valor $\theta_i$ será el mismo para todos los distritos, y por lo tanto, para realizar la estimación en i-ésimo distrito de la característica de interés, se incorpora la información de otros distritos mediante el vector de los coeficientes $\beta$. Se supone que los valores de las varianzas de los errores de muestreo $\sigma_i$ son conocidas. Este supuesto es razonable puesto que varianzas de estimadores directos pueden ser calcu-ladas bajo el diseño del experimento, a partir de los datos muestreados. El modelo (2.1) puede ser reescrito de manera matricial de la siguiente forma:

$$Y = X\beta + u + e, \quad (2.2)$$

donde $Y = (Y_1,...,Y_D)^t$ representa el vector de estimaciones directas calculadas a partir de los datos de la Endes, $u = (u_1,...,u_D)^t \sim N(0,\Sigma_u)$, $e = (e_1,...,e_D)^t \sim N(0,\Sigma_e)$, donde las matrices de covarianzas $\Sigma_u = \sigma_u^2 I_D$ y $[\Sigma_e]_{ij} = \sigma_i^2 I_{(i=j)}$, $i,j = 1, ..., D$ son matrices diagonales. Si el valor de $\sigma_u^2$ es conocido (la situación ideal, puesto que, en la práctica, la varianza poblacional es desconocida), $\theta_i$, $i = 1, ..., D$ puede ser estimado mediante el BLUP de la siguiente forma (véanse Fay & Herriot, 1979; Rao, 2003):

$$\hat{\theta}_i^{\text{BLUP}}(\sigma_u^2) = X_i\hat{\beta}(\sigma_u^2) + \hat{u}_i(\sigma_u^2) \quad (2.3)$$

donde $\hat{\beta}$ se puede escribir como una función de la varianza del efecto aleatorio de la siguiente forma:

$$\hat{\beta}(\sigma_u^2) = (X^t[V(\sigma_u^2)]^{-1}X)^{-1}X^t[V(\sigma_u^2)]^{-1}Y \quad (2.4)$$

y los estimadores de los efectos aleatorios como:

$$\hat{u}_i(\sigma_u^2) = E(u_i | Y_i) = \gamma_i(\sigma_u^2)(Y_i - X_i\hat{\beta}(\sigma_u^2)) \quad (2.5)$$

donde $V(\sigma_u^2) = Var(u + e) = \Sigma_u + \Sigma_e$ y $\gamma_i(\sigma_u^2) = \dfrac{\sigma_u^2}{\sigma_i^2 + \sigma_u^2}$. Entonces, el predictor (2.3) de la característica de interés del presente trabajo puede ser reescrito en función de $\sigma_u^2$ de la siguiente forma:

$$\hat{\theta}_i^{\text{BLUP}}(\sigma_u^2) = \gamma_i(\sigma_u^2)Y_i + (1 - \gamma_i(\sigma_u^2))\sigma_u^2)X_i\hat{\beta}(\sigma_u^2) \quad (2.6)$$

Observen que el predictor (2.6) constituye una combinación convexa del estimador directo $Y_i$ y el estimador basado en el modelo de regresión $X_i\hat{\beta}(\sigma_u^2)$. Si el tamaño de la muestra en el $i$-ésimo distrito es pequeño, la varianza del error de muestreo correspondiente $\sigma_i^2$ será grande, y consecuentemente el valor de $\gamma_i(\sigma_u^2)$, que representa el peso del estimador directo, será pequeño. En esta situación, la predicción para el $i$-ésimo distrito se basa mayormente en el modelo (2.1). Por otro lado, si el tamaño de la muestra en el $i$-ésimo distrito es suficientemente grande, el predictor $\hat{\theta}_i$ se basará mayormente en el estimador directo. Obviamente, si el $i$-ésimo distrito no está representado en la muestra, el valor correspondiente de $\gamma_i$ es igual a cero, y la predicción para este distrito será igual a $X_i\hat{\beta}(\sigma_u^2)$. Por lo tanto, el predictor (2.6) tiene varianza menor comparando con el estimador directo.

¿Es posible obtener estimaciones confiables del porcentaje de anemia y retraso en el crecimiento en niños menores de cinco años en los distritos más pobres del Perú?En la realidad, el valor del parámetro $\sigma_u^2$ es desconocido en la práctica y, generalmente, se estima utilizando el método de máxima verosimilitud (ML), el método de máxima verosimilitud restringida (REML), el método de los momentos, desarrollado en Prasad y Rao (1990), o el estimador propuesto en Fay y Herriot (1979).

Para estimar los parámetros de la característica de interés de la población (porcentaje de anemia y riesgo de retraso en el crecimiento en niños menores de cinco años), definimos las siguientes funciones: el logaritmo de la función de verosimilitud (2.7) y el logaritmo de la función de verosimilitud restringida (2.8). Estas funciones son:

$$l_{ML}(\beta,\sigma_u^2) = c - \tfrac{1}{2}log|V| - \tfrac{1}{2}(Y - X\beta)V^{-1}(Y - X\beta)^t \quad (2.7)$$

donde $c$ es una constante que no influenciara en nuestros cálculos y resultados.

$$l_{REML}(\sigma_u^2) = c - \tfrac{1}{2}log|V| - \tfrac{1}{2}log|X^tV^{-1}X| - \tfrac{1}{2}Y^tPY \quad (2.8)$$

donde $P$ es la siguiente matriz $P = V^{-1} - V^{-1}X(X^tV^{-1}X)^{-1}X^tV^{-1}$.

Contrariamente al método de ML, el método REML toma en cuenta la pérdida de grados de libertad que ocurre por causa de la estimación de los parámetros $\beta$ en el modelo, y, por lo tanto, tiene más ventaja y es más recomendable usarlo en el caso de que el tamaño de la muestra sea pequeño (para una discusión más detallada sobre estimación en áreas pequeñas, el lector puede revisar Rao (2003), Molina *et al.* (2009) y Rao y Molina (2015). El estimador por el método de los momentos para $\sigma_u^2$ obtener se obtiene de la siguiente forma.

$$\tilde{\sigma}_u^2 = \tfrac{1}{D-p} \sum_{i=1}^{D}[(Y_i - X_i\hat{\beta}_{MCO})^2 - \sigma_u^2(1 - h_i)] \quad (2.9)$$

donde $\hat{\beta}_{MCO} = (X^tX)^{-1}X^tY$, $h_i = X_i(X^tX)^{-1}X_i^t$ y $p$ es el número de variables auxiliares en el modelo (2.1). Puesto que el valor de $\tilde{\sigma}_u^2$ en la ecuación (2.9) puede ser negativo, el estimador para $\sigma_u^2$ será igual a:

$$\hat{\sigma}_u^2 = máx\{0, \tilde{\sigma}_u^2\} \quad (2.10)$$

El estimador dado en (2.9), propuesto por Fay y Herriot (1979), se obtiene resolviendo de forma iterativa la ecuación

$$\sum_{i=1}^{D}\frac{(Y_i - X_i\hat{\beta}*)^2}{\sigma_u^{*2} + \sigma_i^2} = (D - p) \quad (2.11)$$

183

Es importante resaltar que los estimadores para $\sigma_u^2$ mencionados anteriormente son invariantes por traslaciones, es decir, tienen las siguientes propiedades (para más detalles, véase Kackar & Harville, 1984):

(i) $\hat{\sigma}_u^2(Y) = \hat{\sigma}_u^2(-Y)$

(ii) $\hat{\sigma}_u^2(Y - Xa) = \hat{\sigma}_u^2(Y)$, $\forall\ a \in R^p$ y $\forall\ Y$.

Reemplazando en la ecuación (2.5) el parámetro $\sigma_u^2$ por su estimador consistente $\hat{\sigma}_u^2$, obtenemos el mejor predictor lineal insesgado empírico (EBLUP) para los valores reales de las características de interés $\theta_i$, $i = 1,...,D$ (véase Fay & Herriot, 1979):

$$\hat{\theta}_i^{\text{EBLUP}}(\hat{\sigma}_u^2) = \gamma_i(\hat{\sigma}_u^2)Y_i + (1 - \gamma_i(\hat{\sigma}_u^2))X_i\hat{\beta}(\hat{\sigma}_u^2) \quad (2.12)$$

En Kackar y Harville (1984), los autores demuestran que si $\hat{\sigma}_u^2$ es un estimador invariante por traslaciones, el predictor $\hat{\theta}_i^{\text{EBLUP}}$ obtenido es insesgado para $\theta_i$.

El modelo de Fay-Herriot es una modificación de un modelo factorial, donde el objetivo principal es buscar o caracterizar la variable de interés a través de un conjunto de variables explicativas sin error de medida. Esto se consigue seleccionando variables de los censos nacionales. Vale notar que los datos de la Endes muestran que tanto los porcentajes de anemia como de riesgo de retraso en el crecimiento presentan una alta autocorrelación a lo largo de los distritos (véase la sección 3). Esta propiedad observada en los datos de la Endes es útil para mejorar el modelo de Fay-Herriot básico (Fay & Herriot, 1979) incorporando la autocorrelación espacial que exista entre los distritos y así obtener una mejor estimación de la característica de interés, sobre todo, una mejor estimación en los distritos con pocas observación y en los distritos que no pertenecen a la muestra que selecciona la Endes para su aplicación. La inclusión de correlaciones espaciales en el modelo original de Fay y Herriot fue desarrollada en Pratesi y Salvati (2008). En la siguiente sección, describimos este modelo.

### 2.2.2 Modelo de Fay-Herriot con información espacial

En la figura 2.2, construida a partir de la Endes 2019, observamos que el porcentaje de niños menores de cinco años con riesgo de retraso en el crecimiento tiene un patrón de comportamiento en *cluster*, es decir, distritos vecinos tienen los valores de la característica de interés altamente correlacionados. Este patrón se observa en todos los departamentos del Perú. Por ejemplo, en el caso del departamento de La Libertad, en el mapa izquierdo de la figura 2.2, observamos que los distritos muestreados más próximos a la costa presentan un porcentaje de riesgo de retraso en el crecimiento de entre un 10% y un 25%, mientras que en los distritos más alejados de la costa se observan porcentajes más altos. Además, calculando la autocorrelación espacial del porcentaje de niños con riesgo de retraso en el crecimiento, obtenemos un valor de 0,74. Notemos también que los distritos con menor valor de la característica de interés están próximos a la capital (en la mayoría de casos). En el caso de Arequipa, mapa derecho de la figura 2.2, observamos un patrón similar, en el que distritos más próximos a la capital tienen menores valores de riesgo de retraso en el crecimiento en los niños menores de cinco años. Esto puede atribuirse a un problema de déficit en la infraestructura de los distritos más alejados de la capital o por las principales actividades económicas a las que se dedica cada distrito.

Figura 2.2
Porcentaje de niños menores de cinco años con retraso en el crecimiento (lado izquierdo: La Libertad; lado derecho: Arequipa)

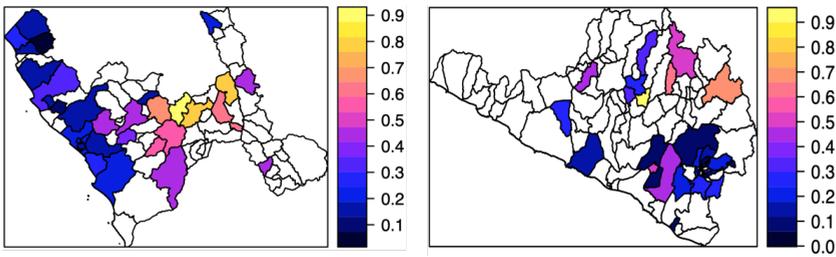

Nota. Los distritos en color blanco no presentan datos.
Fuente: elaboración propia con base en los datos de la Endes 2019.

Entonces, el modelo de Fay-Herriot con información espacial intenta introducir o capturar este patrón observado en los datos de la Endes. Para poder describir la correlación espacial es necesario construir un grafo a partir de los datos, introduciendo alguna definición de vecindad para los distritos en los mapas. A continuación, describimos el modelo.

El modelo de Fay-Herriot con información espacial está definido de la siguiente forma (véase Pratesi & Salvati, 2008):

$$Y = X\beta + u + e; \quad u = \rho W u + \epsilon \quad (2.13)$$

donde $\epsilon = (\epsilon_1,..., \epsilon_D)^t \sim N(0, \Sigma_\epsilon)$ con $\Sigma_\epsilon = \sigma_\epsilon^2 I$, $e$ representa el error del estimador directo, $\rho$ es el coeficiente autorregresivo espacial desconocido (véase Beard & Connor, 2003) y $W$ es la matriz de los pesos espaciales cuyos elementos $w_{ij}$ definen la medida de proximidad espacial entre los distritos $i$ y $j$ en el grafo o mapa. Además, $e$ y $\epsilon$ (error idiosincrático) son independientes, y $\forall\ i = 1,...,D$ tenemos que $w_{ii} = 0$ y $\sum_{j=1}^{D} w_{ij} = 1$ (para más detalles sobre esta condición, véanse Cressie [1993] y Anselin [1992]). Note que el modelo espacial (2.13) introduce un único parámetro $\rho$ para medir la autocorrelación entre los distritos vecinos, y que esta dependencia es lineal en el parámetro $\rho$, es decir, la dependencia espacial entre dos distritos vecinos $i$ y $j$, introducida en el modelo (2.13), está dada por $\rho w_{ij}$, y la influencia total de los distritos vecinos sobre el distrito $i$ sería igual $\rho \sum_{j:i\sim j}$, donde $\sim$ denota que $i$ y $j$ son vecinos, y la suma anterior sería sobre todos los vecinos $j$ del distrito $i$. En la práctica existen los siguientes métodos más utilizados para definir los distritos vecinos (el lector interesado puede encontrar otras definiciones de vecinos comúnmente utilizados en la práctica en Cressie [1993] y Anselin [1992]).

1. Los distritos que están localizados a una distancia menor o igual que $L$ del distrito de interés.
2. Los $k$ distritos más cercanos al distrito de interés. En este caso es posible que un distrito $i$ sea vecino del distrito $j$, y que, sin embargo, el distrito $j$ no sea vecino del distrito $i$.
3. Los distritos limítrofes del distrito de interés, es decir, distritos con frontera geográfica común.

En este trabajo, definimos las entradas $w_{ij}$, $i, j = 1, ..., D$ de la matriz de pesos espaciales $W$ de la siguiente forma:

$$w_{ij} = \frac{1}{K_i} I_{S_i}(j),$$

donde $S_i$ denota el conjunto de los distritos $j$, que son vecinos del distrito $i$ y $K_i = \sum_j I_{S_i}(j)$. Aquí, $I$ denota la función indicadora.

Es importante resaltar que no existe ninguna regla exacta de cómo definir los sitios vecinos de un distrito, esto depende también del tipo de datos que se tiene en la encuesta. Evidentemente, existe una gran cantidad de matrices de pesos espaciales que se pueden utilizar para el mismo modelo espacial. Es importante tener en cuenta que los resultados de cualquier análisis estadístico espacial potencialmente pueden depender de la matriz de pesos espaciales que se elija. Sin embargo, la selección de los pesos espaciales generalmente no afecta significativamente las estimaciones. En este trabajo también verificamos la sensibilidad de las conclusiones para diferentes matrices de pesos $W$, obteniendo resultados muy parecidos.

Para facilitar el análisis y la estimación de los parámetros, el modelo (2.13) puede ser reescrito de la siguiente forma:

$$Y = X\beta + (I - \rho W)^{-1}\epsilon + e = X\beta + v, v \sim N(0, G) \quad (2.14)$$

¿Es posible obtener estimaciones confiables del porcentaje de anemia y retraso en el crecimiento en niños menores de cinco años en los distritos más pobres del Perú?

Donde:

$$G = \sigma_\epsilon^2 [(I - \rho W)^t(I - \rho W)]^{-1} + \Sigma_e = \Omega + \Sigma_e$$

Note que la matriz $G$ solo existe para valores de $\rho$ donde la inversa de la matriz $(I - \rho W)$ existe. Entonces, en el proceso de estimación de los parámetros se tomará en cuenta dicha restricción para la implementación del método de optimización.

Ahora, sean $\phi = (\sigma_\epsilon^2, \rho)$ los parámetros del modelo, y $b_i$ un vector con valor de 1 en la i-ésima posición y 0 en otras posiciones, *i. e.*, $b_i = (0,...,0,1,0,...,0)^t$. Entonces, el predictor BLUP espacial para $\theta_i$, SBLUP se obtiene de manera parecida a como en el modelo anterior:

$$\hat{\theta}_i^{\text{SBLUP}}(\phi) = X_i\hat{\beta}(\phi) + \hat{u}_i(\phi) \quad (2.15)$$

Donde:

$$\hat{\beta}(\phi) = (X^t[G(\phi)]^{-1}X)^{-1}X^t[G(\phi)]^{-1}Y \quad (2.16)$$

y

$$\hat{u}_i(\phi) = b_i^t \Omega^t(\phi)[G(\phi)]^{-1}(Y - X\hat{\beta}(\phi)) \quad (2.17)$$

Los estimadores de los parámetros desconocidos $\phi$ se obtienen aplicando el método ML o REML, donde la matriz de las covarianzas $V(\sigma_\epsilon^2)$ en (2.7) o (2.8) se reemplaza por la matriz $G(\phi)$. Molina *et al.* (2009) advierten sobre posibles problemas numéricos, asociados al proceso de optimización de las funciones definidas en (2.7) y (2.8) en el caso del modelo de Fay-Herriot espacial. La descripción del proceso de optimización de estos modelos puede verse en Rao (2003) y Rao y Molina (2015).

Reemplazando los parámetros $\phi$ por sus estimadores correspondientes en (2.16) y en (2.17) en el predictor SBLUP empírico, obtenemos el predictor SEBLUP para $\theta_i$:

$$\hat{\theta}_i^{\text{SEBLUP}}(\hat{\phi}) = X_i\hat{\beta}(\hat{\phi}) + \hat{u}_i(\hat{\phi}) \quad (2.18)$$

El estimador (2.18) es insesgado para $\theta_i$ si $\hat{\sigma}_\epsilon^2$ y $\hat{\rho}$ son estimadores ML o REML (véase Kackar y Harville, 1984). Estos estimadores serán implementados en el *software* R (Molina & Rao, 2010) para la obtención de las estimaciones de los parámetros para el modelo aplicado a los datos de la Endes.

## 3. Aplicación a los datos de la Endes

### 3.1 Descripción de la encuesta

Como se ha mencionado anteriormente, la Encuesta Demográfica y de Salud Familiar (Endes) es una de las encuestas anuales elaboradas por el Instituto Nacional de Estadística e Informática (INEI) que tiene como objetivo proporcionar información sobre el estado de salud y demográfico principalmente de la población de mujeres de 12 a 49 años de edad y niños menores de cinco años. Entre las principales estadísticas que se derivan de esta encuesta se encuentran la tasa de fecundidad, el uso de métodos de planificación familiar, la prevalencia de anemia en niños, el retraso en el crecimiento, entre otras. La Endes es una encuesta bietápica a nivel nacional, cuyas unidades de muestreo son viviendas. Dicha encuesta se realiza de tal manera que en la primera etapa se seleccionan conglomerados en las áreas urbanas y áreas de empadronamiento rural en las áreas rurales, en todos los departamentos del país con probabilidad proporcional a su tamaño, empleando la información del Censo de Población y Vivienda del 2007 y el Sistema de Focalización de Hogares 2012-2013. En la segunda etapa, se eligieron viviendas de los conglomerados y áreas de empadronamiento rural seleccionados anteriormente, de tal manera que la muestra sea equilibrada considerando las variables: niñas y niños menores de cinco años, mujeres en edad fértil, etc. (véase INEI, 2019). En el año 2019, el número total de viviendas muestreadas fue 36 760.

### 3.2 Descripción de las variables

En este trabajo, vamos a analizar los niveles de prevalencia de anemia y el porcentaje de niños con riesgo de retraso en el crecimiento, a nivel distrital. Según los criterios de la Organización Mundial de la Salud, un niño se considera anémico si su nivel de hemoglobina se encuentra por debajo de 11 g/l; un niño padece de retraso en crecimiento si su talla está en el percentil 2,5 de la curva de crecimiento para su edad. En ambos casos, consideramos los niños menores de cinco años. Para estimar el porcentaje de niños anémicos y de niños con retraso en el crecimiento, por distrito, utilizamos el estimador de Horvitz-Thompson (Horvitz & Thompson, 1992). Para estimar las varianzas de los estimadores, utilizamos las técnicas desarrolladas en Horvitz y Thompson (1952).

Como ya fue mencionado, la fuente utilizada para obtener la información auxiliar es el Censo Nacional del año 2017, basándose en la cual, podemos obtener variables auxiliares sin error de medida relacionadas con el porcentaje a nivel distrital de la población o de las viviendas con ciertas características, tales como: viviendas con acceso a servicios públicos (agua, luz, desagüe), población con acceso a seguro de salud (por ejemplo, SIS, EsSalud, privado), población que tiene como lengua materna el castellano, población con analfabetismo, etc. Además, disponemos de las características geográficas de los distritos: la altitud, longitud y latitud. La altitud del distrito se utilizará como una de las variables auxiliares, y las coordenadas geográficas se emplearán para definir la matriz de los pesos espaciales $W$ que se usará como información para poder estimar mejor las características de interés en distritos no muestreados. En la siguiente tabla, presentamos las variables auxiliares que utilizaremos en este trabajo y que se incluirán en los modelos mencionados.

### 3.3 Modelos

Vamos a ajustar el modelo de Fay-Herriot básico (2.1) y el modelo de Fay-Herriot espacial 2.13 y calcular las predicciones EBLUP y SEBLUP, definidos en 2.12 y 2.18. Para estimar los parámetros desconocidos, utilizamos el método de REML.

Para mejorar el ajuste de los modelos descritos en la sección 2, dividimos los distritos de todos los departamentos en tres estratos, utilizando una variable que indica el porcentaje de la población que vive en pobreza. Después, ajustamos los modelos a cada uno de los estratos separadamente. De acuerdo con el nivel de pobreza, los estratos son los siguientes:

1. Estrato 1: los distritos donde menos del 30% de la población vive en pobreza.
2. Estrato 2: los distritos donde más del 30% y menos del 55% de la población vive en pobreza.
3. Estrato 3: los distritos donde más del 55% de la población vive en pobreza.

Tabla 3.1
Descripción de las variables

| Variable | Descripción de la variable |
| --- | --- |
| Altitud | La altitud del distrito en miles de metros sobre el nivel del mar (en km). |
| PisoTierra | % de viviendas que tienen la tierra como material de construcción del piso de su hogar. |
| Cemento | % de viviendas que tienen el cemento como material de construcción de las paredes de su hogar. |
| Agua | % de viviendas que tienen acceso a agua mediante red pública. |
| Elect. | % de viviendas que tienen acceso a electricidad mediante red pública. |
| Desagüe | % de viviendas que tienen acceso a desagüe mediante red pública. |
| Internet | % de viviendas que tienen conexión a internet. |
| SIS | % de personas que tienen acceso al Seguro Integral de Salud (SIS). |
| EsSalud | % de personas que tienen acceso al Seguro Social de Salud del Perú (EsSalud). |
| NoSeguro | % de personas que no tienen acceso a un seguro de salud. |
| Refrig. | % de hogares que tienen una refrigeradora o congeladora. |
| Castellano | % de personas que tienen como lengua materna el castellano. |
| Analfabet. | % de personas que no saben leer y escribir. |
| Rural | % de viviendas que se encuentran en la zona rural. |

En el primer estrato, se encuentran 618 distritos (234 en la muestra y 384 fuera de la muestra); el segundo estrato incluye 671 distritos (271 en la muestra y 392 fuera de la muestra); y el tercer estrato tiene 585 distritos (260 en la muestra y 325 fuera de la muestra).

Primero, ajustamos los modelos FH y FH con dependencia espacial para explicar la prevalencia de anemia a nivel distrital en cada uno de los estratos definidos anteriormente. En las siguientes tres tablas, presentamos nuestros resultados para la estimación de los coeficientes $\beta$ de las variables auxiliares significativas, donde la variable de interés es el porcentaje de niños anémicos por distrito.

Tabla 3.2
Prevalencia de anemia: estimación de los coeficientes $\beta$ en los distritos con nivel de pobreza menor del 30%

| | FH | | FH espacial | |
|---|---|---|---|---|
| Variable | Coef. | p-val. | Coef. | p-val. |
| Intercepto | 0,494 | 0,000 | 0,472 | 0,000 |
| Altitud | 0,015 | 0,012 | 0,017 | 0,027 |
| PisoTierra | — | — | — | — |
| Cemento | — | — | — | — |
| Agua | — | — | — | — |
| Elect. | — | — | — | — |
| Desagüe | — | — | — | — |
| Refrig. | — | — | — | — |
| Internet | -0,279 | 0,000 | -0,221 | 0,000 |
| SIS | — | — | — | — |
| EsSalud | — | — | — | — |
| NoSeguro | — | — | — | — |
| Castellano | -0,182 | 0,000 | -0,176 | 0,000 |
| Analfabet. | — | — | — | — |
| Rural | — | — | — | — |

¿Es posible obtener estimaciones confiables del porcentaje de anemia y retraso en el crecimiento en niños menores de cinco años en los distritos más pobres del Perú?

Tabla 3.3
Prevalencia de anemia: estimación de los coeficientes β en los distritos con nivel de pobreza entre el 30 y el 55%

|  | F H | | FH espacial | |
|---|---|---|---|---|
| Variable | Coef. | p-val. | Coef. | p-val. |
| Intercepto | 0,834 | 0,000 | 0,752 | 0,000 |
| Altitud | 0,012 | 0,085 | 0,019 | 0,041 |
| PisoTierra | — | — | — | — |
| Cemento | -0,141 | 0,056 | -0,148 | 0,045 |
| Agua | -0,114 | 0,009 | -0,089 | 0,066 |
| Elect. | — | — | — | — |
| Desagüe | — | — | — | — |
| Refrig. | — | — | — | — |
| Internet | -0,327 | 0,001 | -0,369 | 0,008 |
| SIS | -0,277 | 0,001 | -0,189 | 0,039 |
| EsSalud | — | — | — | — |
| NoSeguro | — | — | — | — |
| Castellano | -0,221 | 0,000 | -0,232 | 0,000 |
| Analfabet. | — | — | — | — |
| Rural | -0,054 | 0,143 | -0,071 | 0,056 |

Tabla 3.4
Prevalencia de anemia: estimación de los coeficientes β en los distritos con nivel de pobreza mayor del 55%

|  | FH | | FH espacial | |
|---|---|---|---|---|
| Variable | Coef. | p-val. | Coef. | p-val. |
| Intercepto | 0,795 | 0,000 | 0,772 | 0,000 |
| Altitud | 0,021 | 0,025 | — | — |
| PisoTierra | — | — | — | — |
| Cemento | -0,218 | 0,002 | -0,221 | 0,004 |
| Agua | — | — | — | — |
| Elect. | — | — | — | — |
| Desagüe | -0,134 | 0,020 | — | — |
| Refrig. | — | — | — | — |
| Internet | — | — | — | — |
| SIS | -0,350 | 0,007 | -0,287 | 0,002 |
| EsSalud | — | — | — | — |
| NoSeguro | — | — | — | — |
| Castellano | -0,251 | 0,000 | -0,221 | 0,000 |
| Analfabet. | — | — | — | — |
| Rural | — | — | — | — |

Los resultados obtenidos indican que, en todos los estratos, las variables que explican prevalencia de anemia en los distritos son: Altitud, Internet y Castellano, donde la variable Altitud tiene un impacto positivo y las variables Internet y Castellano tienen un impacto negativo. Es decir, en los distritos que se encuentran en regiones más altas, el porcentaje de niños anémicos es más alto; en las regiones donde la lengua natal es mayoritariamente castellano y en las regiones donde el acceso a internet es mayor, el porcentaje de niños anémicos es más bajo. Además, en los estratos 2 y 3 (donde 30-55% y más del 55% de la población viven en pobreza, respectivamente), las variables Cemento y SIS tienen efecto negativo. Asimismo, las variables Agua y Rural en el estrato 2, y la variable Desagüe en el estrato 3 presentan impactos negativos.

En las tablas 3.5-3.7, se presentan las estimaciones de los parámetros $\beta$ de las variables auxiliares significativas, en los modelos de Fay-Herriot y Fay-Herriot con dependencia espacial, utilizados para explicar el porcentaje de niños con retraso en el crecimiento

Tabla 3.5
Retraso en el crecimiento: estimación de los coeficientes β en los distritos con nivel de pobreza menor del 30%

| Variable | FH | | FH espacial | |
|---|---|---|---|---|
| | Coef. | p-val. | Coef. | p-val. |
| Intercepto | 0,153 | 0,002 | 0,176 | 0,001 |
| Altitud | 0,012 | 0,000 | 0,012 | 0,124 |
| PisoTierra | — | — | — | — |
| Cemento | — | — | — | — |
| Agua | — | — | — | — |
| Elect. | — | — | — | — |
| Desagüe | — | — | — | — |
| Refrig. | -0,337 | 0,000 | -0,388 | 0,000 |
| Internet | -0,202 | 0,032 | 0,269 | 0,005 |
| SIS | — | — | — | — |
| EsSalud | — | — | — | — |
| NoSeguro | — | — | — | — |
| Castellano | — | — | — | — |
| Analfabet. | 1,612 | 0,000 | 1,498 | 0,000 |
| Rural | — | — | — | — |

¿Es posible obtener estimaciones confiables del porcentaje de anemia y retraso en el crecimiento en niños menores de cinco años en los distritos más pobres del Perú?

Tabla 3.6
Retraso en el crecimiento: estimación de los coeficientes β en los distritos con nivel de pobreza entre el 30 el y 55%

| Variable | FH | | FH espacial | |
| --- | --- | --- | --- | --- |
| | Coef. | p-val. | Coef. | p-val. |
| Intercepto | 0,516 | 0,000 | 0,511 | 0,000 |
| Altitud | — | — | — | — |
| PisoTierra | — | — | — | — |
| Cemento | — | — | — | — |
| Agua | 0,117 | 0,048 | 0,118 | 0,051 |
| Elect. | -0,355 | 0,001 | -0,146 | 0,003 |
| Desagüe | — | — | — | — |
| Refrig. | -0,175 | 0,002 | -0,202 | 0,001 |
| Internet | — | — | — | — |
| SIS | — | — | — | — |
| EsSalud | — | — | — | — |
| NoSeguro | -0,282 | 0,025 | -0,260 | 0,042 |
| Castellano | — | — | — | — |
| Analfabet | 0,807 | 0,005 | 0,771 | 0,008 |
| Rural | — | — | — | — |

Tabla 3.7
Retraso en el crecimiento: estimación de los coeficientes β en los distritos con nivel de pobreza mayor del 55%

| Variable | FH | | FH espacial | |
| --- | --- | --- | --- | --- |
| | Coef. | p-val. | Coef. | p-val. |
| Intercepto | 0,611 | 0,000 | 0,647 | 0,000 |
| Altitud | 0,022 | | 0,018 | 0,133 |
| PisoTierra | — | — | — | — |
| Cemento | — | — | — | — |
| Agua | — | — | — | — |
| Elect. | -0,306 | 0,000 | -0,322 | 0,000 |
| Desagüe | — | — | — | — |
| Refrig. | -0,340 | 0,000 | -0,354 | 0,000 |
| Internet | — | — | — | — |
| SIS | -0,326 | 0,002 | -0,345 | 0,002 |
| EsSalud | — | — | — | — |
| NoSeguro | — | — | — | — |
| Castellano | — | — | — | — |
| Analfabet. | 1,486 | 0,000 | 1,492 | 0,000 |
| Rural | — | — | — | — |

Como se observa en las tablas, en todos los estratos, el nivel de analfabetismo en el distrito es significativo, lo cual era de esperarse debido a que los niveles de analfabetismo más altos corresponden a distritos con problemas sociales más graves. Asimismo, otra variable significativa en los tres estratos es la variable Refrig., que puede servir como un indicador de pobreza, y cuyo impacto negativo implica que, en los distritos más pobres, el porcentaje de niños con riesgo de retraso en el crecimiento es más alto. Las otras variables significativas al menos en uno de los estratos son: Agua, Elect., SIS, NoSeguro. La variable Agua tiene un impacto positivo, mientras que el resto tiene un impacto negativo. Los resultados presentados en las tablas 3.2-3.7 indican que no existen diferencias considerables entre las estimaciones con el modelo Fay-Herriot y Fay-Herriot espacial. Sin embargo, en las tablas 3.9-3.12 se observa que, para algunos distritos, las diferencias entre las predicciones bajo ambos modelos son considerables. Por lo tanto, nosotros recomendamos utilizar el modelo de Fay-Herriot espacial para los fines de predicción en distritos no muestreados o con pocas observaciones. La razón de esto también se observa en los residuos del modelo de Fay-Herriot básico, donde tenemos autocorrelaciones espaciales significativas. La aplicación del modelo de Fay-Herriot con dependencia espacial a los datos de la Endes resuelve este problema en todos los casos. La siguiente tabla demuestra las estimaciones de la varianza de los residuos para ambos modelos y las estimaciones del parámetro $\rho$ en el modelo de Fay-Herriot espacial. Como se puede observar, en todos los casos el valor de $\hat{\rho}$ es significativo positivo, ya que la varianza de los residuos del modelo de Fay-Herriot espacial es menor que la obtenida aplicando el modelo de Fay-Herriot básico.

Tabla 3.8
Estimación de los parámetros $\sigma_u^2$, $\sigma_\epsilon^2$, $\rho$

| Modelo | FH | FH espacial | |
|---|---|---|---|
| | $\hat{\sigma}_u^2$ | $\hat{\sigma}_\epsilon^2$ | $\hat{\rho}$ |
| Anemia: estrato 1 | 0,00682 | 0,00524 | 0,570 |
| Anemia: estrato 2 | 0,00670 | 0,00451 | 0,694 |
| Anemia: estrato 3 | 0,00812 | 0,00401 | 0,742 |
| Retraso: estrato 1 | 0,00712 | 0,00566 | 0,474 |
| Retraso: estrato 2 | 0,00982 | 0,00948 | 0,235 |
| Retraso: estrato 3 | 0,01098 | 0,01057 | 0,317 |

¿Es posible obtener estimaciones confiables del porcentaje de anemia y retraso en el crecimiento en niños menores de cinco años en los distritos más pobres del Perú?

## 3.4 Predicción de porcentaje de niños anémicos y de niños con retraso en el crecimiento

Primero, vamos a comparar las predicciones de porcentajes de niños anémicos y de niños con riesgo de retraso en el crecimiento por distrito, en los distritos muestreados. Para este fin, utilizamos el estimador directo y las predicciones EBLUP y SEBLUP, definidos en (2.12) y (2.18), respectivamente. A modo de aplicación de los modelos, los siguientes gráficos presentan dichos estimadores solamente para los departamentos de Puno y Junín, para cada uno de los tres estratos de pobreza considerados en este trabajo.

Figura 3.1
Predicción de porcentaje de niños con anemia por distrito, en los distritos con nivel de pobreza menor del 30%

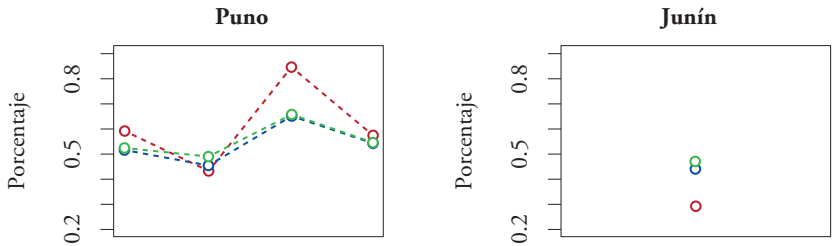

Notas. Estimador directo (rojo), EBLUP (azul) y SEBLUP (verde).

Figura 3.2
Predicción de porcentaje de niños con anemia por distrito, en los distritos con nivel de pobreza entre el 30 y el 55%

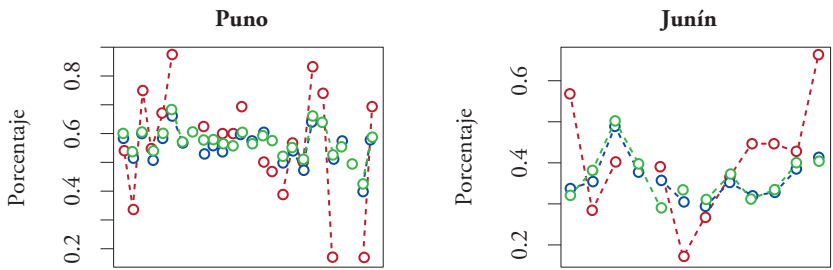

Notas. Estimador directo (rojo), EBLUP (azul) y SEBLUP (verde).

Figura 3.3
Predicción de porcentaje de niños con anemia por distrito, en los distritos con nivel de pobreza mayor del 55%

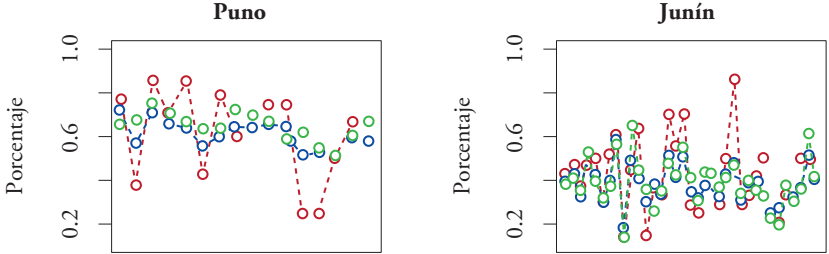

Notas. Estimador directo (rojo), EBLUP (azul) y SEBLUP (verde).

Figura 3.4
Predicción de porcentaje de niños con retraso en el crecimiento por distrito, en los distritos con nivel de pobreza menor del 30%

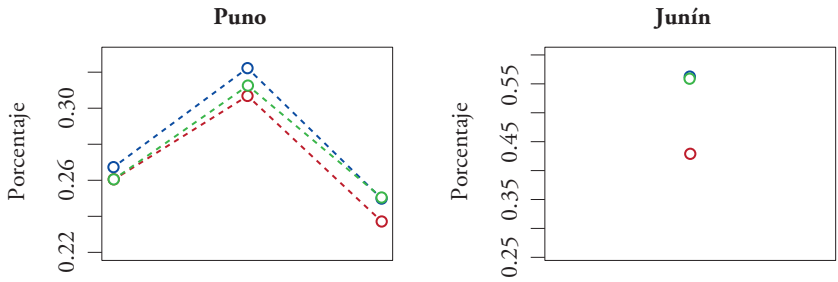

Notas. Estimador directo (rojo), EBLUP (azul) y SEBLUP (verde).

Figura 3.5
Predicción de porcentaje de niños con retraso en el crecimiento por distrito, en los distritos con nivel de pobreza entre el 30 y el 55%

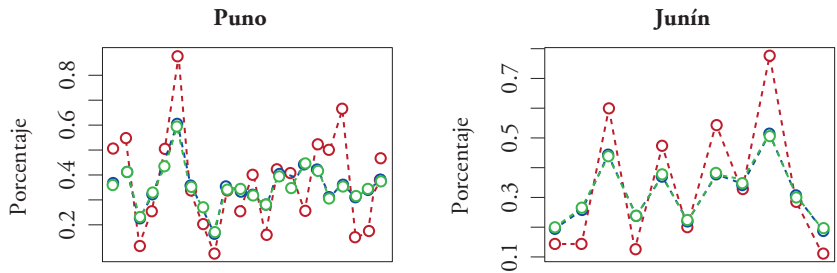

Notas. Estimador directo (rojo), EBLUP (azul) y SEBLUP (verde).

¿Es posible obtener estimaciones confiables del porcentaje de anemia y retraso en el crecimiento en niños menores de cinco años en los distritos más pobres del Perú?

Las figuras 3.1-3.5 indican que las predicciones EBLUP y SEBLUP son generalmente próximas. Asimismo, se puede observar que, comparando con el estimador directo, los predictores EBLUP y SEBLUP reducen considerablemente la variación de las predicciones. La razón de esto es que, para realizar predicciones en un distrito específico, ambos predictores utilizan la información adicional contenida en las variables auxiliares del mismo distrito y la información de los otros distritos mediante los parámetros $\beta$. Además, el predictor SEBLUP utiliza la información de los otros distritos mediante las autocorrelaciones espaciales.

Finalmente, empleando los modelos de Fay-Herriot básico y Fay-Herriot espacial, las siguientes tablas presentan las predicciones del porcentaje de los niños anémicos y de los niños con riesgo de retraso en el crecimiento en algunos distritos no muestreados de los departamentos de Puno y Junín, es decir, usamos los predictores EBLUP y SEBLUP en algunos distritos que no fueron incluidos en la muestra de la Endes.

Figura 3.6
Predicción de porcentaje de niños con retraso en el crecimiento por distrito, en los distritos con nivel de pobreza mayor del 55%

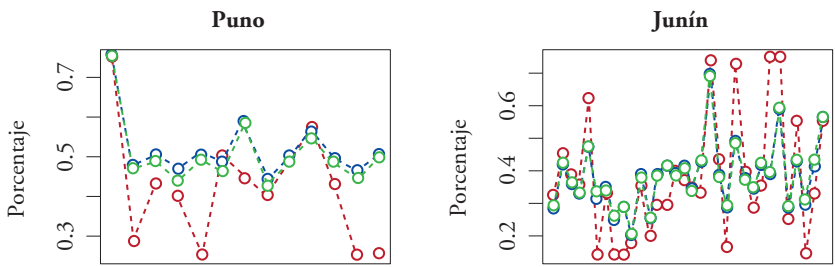

Notas. Estimador directo (rojo), EBLUP (azul) y SEBLUP (verde).

Tabla 3.9
Predicción del porcentaje de niños con anemia por distrito en el departamento de Puno

| Ubigeo | Pobreza (%) | EBLUP | SEBLUP | Ubigeo | Pobreza (%) | EBLUP | SEBLUP |
|---|---|---|---|---|---|---|---|
| 210702 | < 30 | 51,88 | 52,70 | 211105 | < 30 | 44,56 | 44,28 |
| 211203 | < 30 | 53,22 | 53,80 | 210104 | 30-55 | 56,86 | 57,05 |
| 210211 | 30-55 | 54,89 | 56,07 | 210214 | 30-55 | 60,49 | 60,26 |
| 210302 | 30-55 | 54,72 | 55,30 | 210306 | 30-55 | 55,78 | 58,00 |
| 210310 | 30-55 | 63,86 | 63,49 | 210404 | 30-55 | 63,87 | 61,72 |
| 210503 | 30-55 | 57,18 | 56,07 | 210505 | 30-55 | 55,61 | 56,41 |
| 210704 | 30-55 | 59,95 | 58,92 | 210705 | 30-55 | 61,07 | 59,41 |
| 210706 | 30-55 | 55,98 | 57,01 | 210708 | 30-55 | 56,79 | 57,43 |
| 210710 | 30-55 | 58,18 | 58,56 | 210803 | 30-55 | 49,34 | 50,98 |
| 210809 | 30-55 | 55,07 | 54,34 | 211003 | 30-55 | 59,53 | 59,05 |
| 211201 | 30-55 | 48,76 | 48,75 | 211205 | 30-55 | 60,47 | 59,01 |
| 210111 | > 55 | 59,37 | 66,41 | 210202 | > 55 | 63,96 | 73,42 |
| 210305 | > 55 | 55,33 | 66,59 | 210504 | > 55 | 58,68 | 63,55 |
| 210602 | > 55 | 59,66 | 62,96 | 210606 | > 55 | 57,74 | 67,63 |
| 210805 | > 55 | 53,82 | 63,23 | 210901 | > 55 | 60,72 | 62,16 |
| 211004 | > 55 | 53,88 | 64,26 | 211302 | > 55 | 74,16 | 69,11 |

Tabla 3.10
Predicción del porcentaje de niños con anemia por distrito en el departamento de Junín

| Ubigeo | Pobreza (%) | EBLUP | SEBLUP | Ubigeo | Pobreza (%) | EBLUP | SEBLUP |
|---|---|---|---|---|---|---|---|
| 120207 | < 30 | 36,32 | 37,29 | 120908 | < 30 | 36,85 | 36,17 |
| 120214 | 30-55 | 37,53 | 36,98 | 120215 | 30-55 | 25,50 | 25,81 |
| 120407 | 30-55 | 34,52 | 34,13 | 120410 | 30-55 | 25,79 | 24,71 |
| 120411 | 30-55 | 32,14 | 32,54 | 120421 | 30-55 | 30,17 | 31,78 |
| 120428 | 30-55 | 35,64 | 34,59 | 120431 | 30-55 | 29,58 | 29,28 |
| 120502 | 30-55 | 34,80 | 35,63 | 120607 | 30-55 | 36,49 | 36,62 |
| 120106 | > 55 | 43,14 | 52,19 | 120108 | > 55 | 48,53 | 50,20 |
| 120112 | > 55 | 43,55 | 45,04 | 120117 | > 55 | 35,24 | 29,48 |
| 120119 | > 55 | 29,82 | 30,07 | 120122 | > 55 | 40,11 | 36,45 |
| 120126 | > 55 | 41,15 | 36,82 | 120130 | > 55 | 30,37 | 27,69 |
| 120206 | > 55 | 39,57 | 37,43 | 120211 | > 55 | 34,84 | 31,84 |
| 120212 | > 55 | 40,59 | 33,93 | 120306 | > 55 | 34,80 | 31,41 |
| 120409 | > 55 | 33,05 | 33,72 | 120417 | > 55 | 35,27 | 41,66 |
| 120425 | > 55 | 35,41 | 38,03 | 120429 | > 55 | 49,88 | 44,88 |
| 120705 | > 55 | 50,98 | 43,41 | 120803 | > 55 | 39,35 | 42,23 |

¿Es posible obtener estimaciones confiables del porcentaje de anemia y retraso en el crecimiento en niños menores de cinco años en los distritos más pobres del Perú?

Tabla 3.11
Predicción del porcentaje de niños con retraso en el crecimiento por distrito en el departamento de Puno

| Ubigeo | Pobreza (%) | EBLUP | SEBLUP | Ubigeo | Pobreza (%) | EBLUP | SEBLUP |
|---|---|---|---|---|---|---|---|
| 210309 | < 30 | 33,50 | 34,40 | 211203 | < 30 | 41,39 | 41,95 |
| 211105 | < 30 | 35,45 | 36,64 | 210104 | 30-55 | 35,47 | 35,82 |
| 210211 | 30-55 | 39,90 | 40,51 | 210214 | 30-55 | 40,21 | 40,41 |
| 210302 | 30-55 | 49,48 | 49,53 | 210306 | 30-55 | 42,29 | 42,80 |
| 210310 | 30-55 | 44,28 | 44,81 | 210404 | 30-55 | 26,68 | 27,74 |
| 210406 | 30-55 | 37,26 | 37,94 | 210505 | 30-55 | 50,00 | 49,92 |
| 210704 | 30-55 | 33,16 | 33,81 | 210705 | 38,54 | 39,00 | 59,41 |
| 210706 | 30-55 | 53,25 | 52,98 | 210708 | 30-55 | 40,78 | 41,12 |
| 210710 | 30-55 | 46,36 | 46,50 | 210803 | 30-55 | 40,62 | 41,12 |
| 210809 | 30-55 | 31,09 | 31,92 | 211003 | 30-55 | 45,13 | 45,10 |
| 211103 | 30-55 | 34,58 | 35,33 | 211205 | 30-55 | 42,11 | 42,37 |
| 210111 | > 55 | 56,56 | 57,61 | 210202 | > 55 | 48,92 | 49,74 |
| 210305 | > 55 | 53,06 | 52,98 | 210504 | > 55 | 53,64 | 55,08 |
| 210602 | > 55 | 55,00 | 56,24 | 210606 | > 55 | 45,27 | 44,52 |
| 210805 | > 55 | 43,43 | 43,65 | 210901 | > 55 | 46,31 | 46,15 |
| 211004 | > 55 | 49,83 | 50,12 | 211302 | > 55 | 43,93 | 45,50 |

Tabla 3.12
Predicción del porcentaje de niños con retraso en el crecimiento por distrito en el departamento de Junín

| Ubigeo | Pobreza (%) | EBLUP | SEBLUP | Ubigeo | Pobreza (%) | EBLUP | SEBLUP |
|---|---|---|---|---|---|---|---|
| 120207 | < 30 | 42,07 | 42,47 | 120908 | < 30 | 38,66 | 39,18 |
| 120214 | 30-55 | 31,15 | 32,25 | 120215 | 30-55 | 34,34 | 34,91 |
| 120407 | 30-55 | 28,64 | 29,44 | 120410 | 30-55 | 34,01 | 34,88 |
| 120411 | 30-55 | 42,03 | 42,49 | 120413 | 30-55 | 29,24 | 30,11 |
| 120428 | 30-55 | 30,28 | 31,19 | 120431 | 30-55 | 33,33 | 34,08 |
| 120502 | 30-55 | 35,10 | 35,98 | 120607 | 30-55 | 39,88 | 39,92 |
| 120106 | > 55 | 44,28 | 46,09 | 120108 | > 55 | 45,28 | 46,54 |
| 120112 | > 55 | 45,38 | 45,57 | 120117 | > 55 | 34,82 | 35,71 |
| 120119 | > 55 | 32,38 | 34,21 | 120122 | > 55 | 40,73 | 41,80 |
| 120126 | > 55 | 37,87 | 37,71 | 120130 | > 55 | 31,01 | 31,80 |
| 120206 | > 55 | 49,56 | 49,36 | 120211 | > 55 | 30,64 | 31,01 |
| 120212 | > 55 | 38,24 | 39,23 | 120306 | > 55 | 32,76 | 33,03 |
| 120409 | > 55 | 33,74 | 34,68 | 120417 | > 55 | 40,17 | 40,27 |
| 120425 | > 55 | 31,69 | 32,04 | 120429 | > 55 | 46,34 | 47,07 |
| 120705 | > 55 | 36,68 | 37,66 | 120803 | > 55 | 39,09 | 42,18 |

## 4. Conclusiones

En este trabajo, hemos especificado y estimado modelos autorregresivos espaciales para estimar el porcentaje de niños anémicos y niños con riesgo de retraso en el crecimiento por distrito del Perú donde se tiene pocas observaciones, y para predecir dichos porcentajes en distritos donde no se tiene observaciones. Dicho problema es recurrente en muchas encuestas socioeconómicas, puesto que el objetivo de estas es investigar el estado de la población a nivel nacional, y no a niveles más desagregados. Nosotros demostramos que, para compensar las varianzas grandes de los estimadores directos que corresponden a las áreas con poca información (distritos), se puede utilizar la información auxiliar de las mismas áreas y de las áreas vecinas, mediante los modelos de Fay-Herriot básico y Fay-Herriot espacial. El uso de dichos modelos resulta en predicciones más confiables que los estimadores directos. Además, utilizando los modelos mencionados es posible realizar predicciones en las áreas que no fueron incluidas en la muestra.

En este artículo, hemos aplicado la metodología presentada para predecir el porcentaje de niños anémicos y niños con riesgo de retraso en el crecimiento por distrito en el Perú. Los resultados indican que la información auxiliar del censo nacional es relevante para predecir dichos porcentajes. Evidentemente, a los distritos con el nivel de pobreza más alto corresponden porcentajes más altos de niños anémicos y niños con retraso de crecimiento. Específicamente, las variables como el nivel de analfabetismo, el porcentaje de personas que tienen como lengua materna el castellano, el porcentaje de hogares que tienen una refrigeradora o congeladora, etc., han salido significativas en los modelos ajustados. Asimismo, se logró mostrar que la información sobre los distritos vecinos utilizada por el modelo de Fay-Herriot espacial reduce las varianzas de los errores del modelo, lo cual era de esperarse puesto que el análisis preliminar de los datos reveló autocorrelaciones espaciales considerables. A raíz de lo dicho, la conclusión general de la investigación presentada es que el uso del modelo de Fay-Herriot básico y el de Fay-Herriot espacial puede mejorar considerablemente las predicciones en áreas que no cuentan con muestras suficientes para obtener predicciones confiables.

¿Es posible obtener estimaciones confiables del porcentaje de anemia y retraso en el crecimiento en niños menores de cinco años en los distritos más pobres del Perú?

**Referencias**